\documentclass[manuscript]{acmart}
\AtBeginDocument{%
  \providecommand\BibTeX{{%
    \normalfont B\kern-0.5em{\scshape i\kern-0.25em b}\kern-0.8em\TeX}}}

\usepackage{xcolor}

\usepackage{algpseudocode}
\usepackage{algorithm}
\usepackage{bbm, dsfont}

\usepackage{graphicx}

\setcopyright{acmcopyright}
\copyrightyear{2023}
\acmYear{2023}
\acmDOI{XXXXXXX.XXXXXXX}

\acmConference[CONSEQUENCES - RecSys '23]{CONSEQUENCES - RecSys '23}{September 18th-22nd,
  2023}{Singapore}
\begin{document}

\title{Unbiased Offline Evaluation for Learning to Rank with Business Rules}

\author{Matej Jakimov}
\email{jakimate@amazon.de}
\orcid{0009-0006-5619-4862}
\affiliation{%
  \institution{Amazon Music}
  \country{Germany}
}
\author{Alexander Buchholz}
\email{buchhola@amazon.de}
\orcid{0000-0001-6521-7583}
\affiliation{%
  \institution{Amazon Music}
  \country{Germany}
}
\author{Yannik Stein}
\email{syannik@amazon.de}
\affiliation{%
  \institution{Amazon Music}
  \country{Germany}
}
\author{Thorsten Joachims}
\email{tj@cs.cornell.edu}
\orcid{0000-0003-3654-3683}
\affiliation{%
  \institution{Cornell University}
  \country{USA}
}

\renewcommand{\shortauthors}{Jakimov et al.}

\begin{abstract}
  For industrial learning-to-rank (LTR) systems, it is common that the output of a ranking model is modified, either as a results of post-processing logic that enforces business requirements, or as a result of unforeseen design flaws or bugs present in real-world production systems. This poses a challenge for deploying off-policy learning and evaluation methods, as these often rely on the assumption that rankings implied by the model’s scores coincide with displayed items to the users.  
  Further requirements for reliable offline evaluation are proper randomization and correct estimation of the propensities of displaying each item in any given position of the ranking, which are also impacted by the aforementioned post-processing. 
  We investigate empirically how these scenarios impair off-policy evaluation for learning-to-rank models. We then propose a novel correction method based on the Birkhoff-von-Neumann decomposition that is robust to this type of post-processing. We obtain more accurate off-policy estimates in offline experiments, overcoming the problem of post-processed rankings. To the best of our knowledge this is the first study on the impact of real-world business rules on offline evaluation of LTR models.
\end{abstract}




\maketitle

\section{Introduction}
A/B testing is the gold standard for evaluating learning-to-rank (LTR) systems in industrial recommender systems. However, reliable offline evaluation of new ranking policies is often even more important as it allows researchers to evaluate a large number of possible ranking policies without deployment to production. By finding the most promising ranking policies offline, the risk of suboptimal user experience is reduced.

Off-policy evaluation of new policies on historic data requires adequate strategies to deal with biases coming from the way users interact with the system, i.e., commonly referred to as \emph{click models} \citep{chuklin2022click}.
Previous work showed that estimators leveraging randomization of rankings such as in the Item-Position Model~\cite{IPM} (IPM) can be more accurate compared to methods assuming a stronger model of user behavior such as the Position-Bias Model (PBM) \citep{PB, INTERPOL}. If only a subset of candidates is presented to users, randomization is necessary to avoid selection bias \cite{stochasticPBM}. 

Typical architectures of industrial LTR system consist of at least four layers: candidate generation, ranking, post-processing and presentation layer. Many LTR systems in industry \cite{recombee} use some kind of business rules that intentionally modify the ranking in the post-processing step. For example, marketing considerations (or sponsored content) might justify to increase exposure of certain items by displaying them in a top position.
In case randomization and propensity computation is performed in the ranking layer, the propensities no longer represent the actual probabilities of presenting an item at some position. This propensity distortion can bias off-policy estimation.


\section{Background}
\label{sec:background}
Off-policy evaluation methods allow us to estimate a reward $r$ (i.e., number of clicks, conversions, Discounted Cumulative Gain, etc.) for a new policy $\pi$, called \emph{target policy}, provided observations $(x_i, y_i, \{r_{i,1}^{\pi_0}, \dots, r_{i,n}^{\pi_0}\})$, indexed by observation $i$, generated by another policy $\pi_0$, called \emph{logging policy}. As the space of possible rankings $y_i \sim \pi(x_i)$ of length $n$ is usually very large ($O(n!)$), a common assumption is that the reward $r_i$ for the whole ranking $y_i$ can be decomposed into a sum of rewards $r_{i, j}$, one for each item $j$ in the ranking \cite{DeepProp}. We also assume that the logging policy ensures a non-zero probability of observing a reward for every item for which the target policy has a non-zero probability of observing a reward. This assumption is called the \emph{full-support assumption}. We then obtain an unbiased estimate of the reward $r_i$ for $\pi$ using observations $(x_i, y_i, \{r_{i,1}^{\pi_0}, \dots, r_{i,n}^{\pi_0}\})$ generated by logging policy $\pi_0$:

\begin{align}
    \mathbb{E}_\pi[r_i] = \sum_{j=1}^{n} w(y_{i,j}) \cdot \lambda(j) \cdot r_{i,j}^{\pi_0}
    \label{eq:ips}
\end{align}

where $r_{i,j}^{\pi_0}$ is the reward for item $y_{i,j}$ (e.g. $1$ if clicked), $w(\cdot)$ is a weight determined by the used estimator (e.g. PBM, IPM \citep{IPM} or INTERPOL \citep{INTERPOL}) and $\lambda(\cdot)$ is a function of displayed position that defines the metric of interest (e.g. $\lambda(j) = log(1 + j)^{-1}$ for DCG, see \cite{DeepProp}).

For the \textbf{PBM} we define $w_{PBM}(y_j)$ as a ratio of position-biases $b$ for positions where the target ($rank^{\pi}$) and logging policy ($rank^{\pi_0}$) placed the item $y_j$: $w_{PBM}(y_j) = \frac{b_{rank^{\pi}(y_j)}}{b_{rank^{\pi_0}(y_j)}}$. The set of position-biases, one for each position, is called position-bias curve.
It is possible to estimate the position-bias curve from logged data using Expectation-Maximization \citep{wang2018position}, but estimating position-bias curve from randomized data tends to be more accurate \cite{InterventionHarvesting, ruffini2022}. While the PBM estimator usually provides estimates with low variance, it is sensitive to imperfect estimation of the position-bias curve which can lead to bias in the estimated reward \cite{INTERPOL}.

The \textbf{IPM} \cite{IPM} defines \footnote{For simplicity we assume that the target policy is deterministic.} $w_{IPM}(y_j) = \frac{\mathbbm{1}(rank^{\pi_0}(y_j) = rank^{\pi}(y_j)}{\mathbb{P}_{\pi_0}(rank^{\pi_0}(y_j) = rank^{\pi}(y_j))}$. This model thus requires non-zero probability of an item being displayed at any position by $\pi_0$. Typically, the IPM is less prone to bias at the cost of increased variance~\citep{IPM}.

\textbf{INTERPOL} \cite{INTERPOL} interpolates between the PBM and IPM model by using the PBM for local \emph{windows} of neighboring items, whereas across windows the IPM model is used. This allows to trade-off some bias for variance and minimize the MSE of the offline evaluation compared to the plain PBM or IPM estimators.

\paragraph{Randomization Schemes and Birkhoff-von-Neumann Decomposition}
The estimators introduced above require $\pi_0$ to be stochastic and assume knowledge of probabilities of placing items $j$ at position $k$ in the form of a doubly-stochastic propensity matrix $P_{j,k}$. There are several options for obtaining suitable logging policies $\pi_0$ that respect the full-support assumption $P_{j,k} > 0$. One is to use a LTR policy that is inherently stochastic, such as a policy based on nested softmax sampling \cite{PL,QuasiMC}. However, computing propensities requires computationally expensive Monte Carlo estimation via sampling of possible rankings. Another option is to use a deterministic ranker and apply a randomization scheme to the resulting ranking, such as randomization based on Birkhoff-von-Neumann decompositions (BvN) \cite{BvN-fairness, BvN-decomp}.

The BvN scheme decomposes a predetermined propensity matrix $P$ into a set of permutation matrices $\Pi_1, \dots, \Pi_M$ that can be applied to deterministic rankings, each with some probability $p_1, \dots, p_M$, so that: $P = \sum_{m = 1}^{M}p_m \Pi_m$ and $\sum_{m = 1}^{M}p_m =1$. Let $n$ be the number of available items for ranking, then the BvN decomposition of size at most $n^2$ can be computed with time complexity $O(n^{4.5})$. See \citep{BvN-decomp, Hopcroft1973} for more details.

\vspace{-0.1cm}
\paragraph{Business Rules and Other Modifications to Ranking}
In practice the ranking provided by LTR algorithm is often post-processed, both intentionally, in a form of \emph{business rules}, and unintentionally due to unexpected behavior in the processing pipeline (i.e., software bugs). 

One of the most common type of business rules is enforcement of visibility for certain items by always displaying them at the top positions, i.e. \emph{pinning}. Another common type of intentional post-processing is diversification \citep{kunaver2017diversity}. 

As an example of unintentional modification to ranking, imagine a system that only presents an item to the user if the system successfully fetches the corresponding image. If the item is newly added to the system, it is possible that the image is not yet present in a cache and the request for the image times out. We only consider \emph{pinning} in the rest of the paper.

\vspace{-0.2cm}
\section{Proposed Solution}
If the ranking is post-processed between the randomization step and the presentation layer, the propensity matrix $P_{j,k}$ of the randomization scheme no longer represents the true probabilities of an item $j$ being displayed at position $k$. This can lead to a biased estimate of the reward. We assume that it is known what business rules were applied and how they would have been applied to alternative rankings. Then the propensity matrix $P_{j,k}$ can be corrected and a new propensity matrix $P_{j,k}'$, describing the actual propensities after application of business rules, can be derived. In general we could use Monte Carlo estimation of $P^{\prime}$ (see Algorithm \ref{alg:mc} from Appendix \ref{sec:variants}), however such approach is computationally expensive and introduces another source of variance in the off-policy evaluation.

If the BvN randomization scheme is used then it is possible to obtain an exact $P'$ with complexity $O(nM)$, where $n$ is the number of ranked items and $M$ is the size of the BvN decomposition. We represent a ranking $y$ by a matrix $Y$ where $Y_{j,k} = 1$ if item $j$ was ranked at position $k$ and $0$ otherwise. We assume that the deterministic part of the logging policy $\pi_0^{\text{ranker}}$ ranks items by their index, i.e. $Y = I$. We define a function $B(Y)$ that returns a permutation matrix corresponding to the application of all business rules to the ranking $Y$. To obtain $P^{\prime}$, we iterate over all permutations from the BvN decomposition that the logging policy used, and apply the business rules to the permuted rankings: $P^{\prime} = \sum_{m=1}^{M} B(\Pi_m) \cdot p_m \Pi_m$, as explained in Algorithm \ref{alg:main}. Thereby we recover the corrected propensity matrix $P^{\prime}$. 

\begin{algorithm}
\caption{Estimation using BvN}
\label{alg:main}
    \begin{algorithmic}
    \Require observation $x_i$, business rules $B$, BvN decomposition from logging policy $\Pi_1, \dots, \Pi_M, p_1, \dots, p_M$ and the deterministic ranker from logging policy $\pi^{\text{ranker}}_0$
        \State $P^{\prime}_{j,k} \gets 0$
        \State $Y \gets \pi^{\text{ranker}}_0(x_i)$ \Comment{Ranking for the observation $x_i$ provided by deterministic ranker}
        \For{$m \in [1 \dots M]$} \Comment{Iterating over the BvN decomposition used by the logging policy}
            \State $Y^m \gets$  $\Pi_m \cdot Y$ \Comment{Permute the ranking by $\Pi_m$}
            \State $Y^{\prime} \gets$ $B(Y^m)Y^m$ \Comment{Permute the ranking $Y^m$ by applying the business rules}
            \State $P^{\prime} \gets P^{\prime} + p_m Y^{\prime}$ \Comment{$\pi_0$ could have applied $\Pi_m$ with probability $p_m$, add $p_m$ for each item $j$ and its final position}
        \EndFor
    \end{algorithmic}
\end{algorithm}

\paragraph{Deterministic vs. Stochastic Business Rules}
In case business rules are applied with probability $1$, corrected propensity matrix can contain values $P_{j,k}^{\prime} \in \{0,1\}$ and violate the full-support assumption from Section~\ref{sec:background}. This can be avoided by ensuring that modifications to the ranking are only applied with some probability. This is the case for many unintentional modifications due to flaws in the system. For business rules it is often acceptable to apply them only with high probability. An algorithm that incorporates stochasticity of business rules can be found in the Appendix, see Algorithm \ref{alg:stochastic} in Appendix \ref{sec:variants}. This algorithm generalizes Algorithm \ref{alg:main} and will be used in our experiments. 

\section{Experiments}
\begin{figure}
    \centering
    \includegraphics[width=0.9\textwidth]{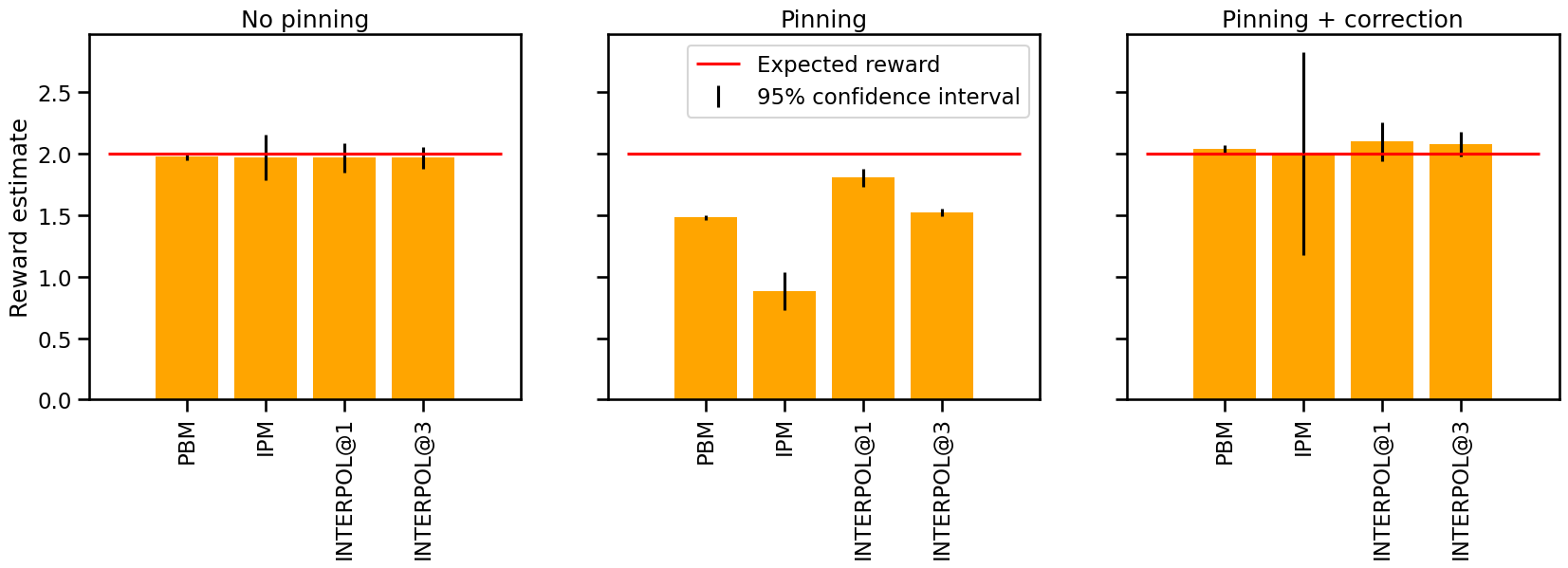}
    \caption{We show the effect of pinning an item to the first position on accuracy of off-policy estimators: PBM (with position-bias curve estimated by \emph{PA-IH} algorithm from \citep{ruffini2022}), IPM and INTERPOL with \emph{window size} set to $1$ and $3$ respectively. In the middle graph, the ranking is first randomized and propensities are estimated. Then one of the items with low relevance is moved to the first position with 95\% probability, which invalidates the propensities. In the right graph we apply pinning with 95\% probability and then post-process propensity matrix using Algorithm \ref{alg:main}.}
    \label{fig:main-results}
\end{figure}

To evaluate the effectiveness of our solution we investigate the effect of pinning an item to the first position in the logging policy. We follow the setup of \cite{INTERPOL}. A full description is given in Appendix \ref{sec:simulations}. The left graph in the Figure \ref{fig:main-results} shows accurate estimates when no pinning is applied, as the expected reward (red line) of the target policy is within the 95\% confidence interval for the mean of the reward. In the middle graph the pinning is applied. As expected, this biases the results. When propensities are corrected by using Algorithm \ref{alg:stochastic}, the results are more accurate and we recover the expected reward. More detailed results are in Appendix \ref{sec:additional-results}.

\section{Conclusions and Future Work}
We show that a randomization scheme based on the Birkhoff-von-Neumann decomposition enables the correction of otherwise biased propensity matrices.  This allows practitioners to achieve unbiaed off-policy evaluation when the ranking was post-processed by business rules, as often occurs in real-world production systems. We also show that such post-processing can considerably bias off-policy estimation if no correction is made. To the best of our knowledge, this is the first contribution investigating the effect of business rules (or other modifications to the ranking) on the quality of off-policy evaluation. In the future we will investigate additional types of modifications, such as dropping certain items, diversification and scenarios where not all ranked items are presented to the user (as in the limited visibility setting \cite{stochasticPBM}). We also plan to study in more detail how modifications to the ranking impact accuracy of off-policy evaluation and off-policy learning in the real-world setting, using data from a deployed production system.

\bibliographystyle{ACM-Reference-Format}
\bibliography{./references.bib}

\appendix

\section{Related Work}
The impact of business rules and other ranking modifications on off-policy evaluation and learning has been acknowledged in recent work, see for example \cite{recombee,London2022} and hence creates a clear need for addressing this issues in real world production systems. \cite{London2022} approaches this problem by detecting potential issues that come from biased propensities. Our work provides a practical solution to de-bias propensities with the aim of enabling unbiased offline evaluation. 

Unbiased offline evaluation has been carefully studied over recent years (see for example \cite{joachims2017unbiased, ai2018unbiased}), but mostly focuses on biases coming from user behavior, not from intentional modifications of the ranking output as our work does.

We build on work investigating the use of randomization of rankings based on Birkhoff-von-Neumann decompositions, see \cite{BvN-fairness} and the use of propensities for unbiased off-policy evaluation as in \cite{INTERPOL, ruffini2022, IPM, DeepProp}.

\section{Simulations}
\label{sec:simulations}
We generated dataset with 50,000 rankings using following scheme:
\begin{enumerate}
    \item each ranking has 10 rankable items encoded as one-hot vectors $u_j$ where $j \in \{0, \dots, 9\}$
    \item random standard normal noise is added to all vectors $u_j$
    \item for each item, relevance vector $v$ is generated where items $\{v_1,v_2,v_4,v_7\}$ have all dimensions set to $1$ and rest of items have all dimensions set to $-1$
    \item items are ranked by score computed as $u_j \cdot v_j$
    \item ranking is randomized by using a BvN decomposition of propensity matrix which keeps the item on the position assigned by ranker with probability $0.95$ and with probability $\frac{0.05}{9}$ it's placed on any other position
    \item optionally, depending on the experiment, some item is pinned/moved to another position
    \item optionally, depending on the experiment, the propensity matrix is corrected using algorithm \ref{alg:main} or \ref{alg:stochastic}
    \item clicks are simulated assuming a PBM with position-bias inversely proportional to the item position $k$: $\frac{1}{k}$ and relevance defined as $\mathbbm{1}_{u_j \cdot v_j > 0}$
\end{enumerate}
Then we evaluated a policy that assigns items $[7, 0, 3, 1]$ to the top and items $[2, 4]$ to the bottom of the ranking in the given order. Rest of the items are ranked arbitrarily. Expected reward (number of clicks per ranking) is known and equal to $2$. The position-bias curve was estimated using \emph{PA-IH} algorithm from \cite{ruffini2022} using an SGD optimizer with decreasing learning rate. Results are shown in Figure \ref{fig:additional-curves} in Appendix \ref{sec:additional-results}.

\section{Variants of the algorithm}
\label{sec:variants}
\begin{algorithm}
    \caption{Monte Carlo Estimation}\label{alg:mc}
    \begin{algorithmic}
        \Require observation $x_i$, business rules $B$ and the ranker from logging policy $\pi^{\text{ranker}}_0$
        \State $P^{\prime}_{j,k} \gets 0$
        \For{L samples}
            \State $Y \gets \pi^{\text{ranker}}_0(x_i)$ \Comment{Sample ranking $Y$ for observation $x_i$ provided by stochastic or randomized ranker}
            \State $Y^{\prime} \gets B(Y) \cdot Y$ \Comment{Apply business rules to ranking $Y$}
            \State $P^{\prime} \gets P^{\prime} + \frac{1}{L} Y^{\prime}$ \Comment{Sum occurrences of the items at their positions after applying business rules}
        \EndFor
    \end{algorithmic}
\end{algorithm}

\begin{algorithm}
    \caption{Estimation using BvN with stochastic business rules}\label{alg:stochastic}
    \begin{algorithmic}
        \Require observation $x_i$, set of applicable business rules $B$, BvN decomposition from logging policy $\Pi_1, \dots, \Pi_M, p_1, \dots, p_M$ and the deterministic ranker from logging policy $\pi^{\text{ranker}}_0$
        \State $P^{\prime}_{j,k} \gets 0$
        \State $Y \gets \pi^{\text{ranker}}_0(x_i)$ \Comment{Ranking for the observation $x_i$ provided by deterministic ranker}
        \For{$m \in [1 \dots M]$} \Comment{Iterating over the BvN decomposition used by the logging policy}
            \State $Y^m \gets$  $\Pi_m \cdot Y$ \Comment{Permute the ranking by $\Pi_m$}
            \For{$S \in \mathcal{P}(B)$} \Comment{where $\mathcal{P}(B)$ is the power set of all applicable business rules $B$. }
                \State $Y^{\prime} \gets$ $S(Y^m)Y^m$ \Comment{Permute the ranking $Y^m$ by applying the subset of business rules}
                \State $P^{\prime} \gets P^{\prime} + p_m \mathbb{P}(S|x_i) Y^{\prime}$ \Comment{Where $\mathbb{P}(S)$ is a probability of applying all business rules from $S$ simultaneously}
            \EndFor
        \EndFor
    \end{algorithmic}
\end{algorithm}

\newpage

\section{Additional Results}
\label{sec:additional-results}
\begin{figure}
    \centering
    \includegraphics[width=\textwidth]{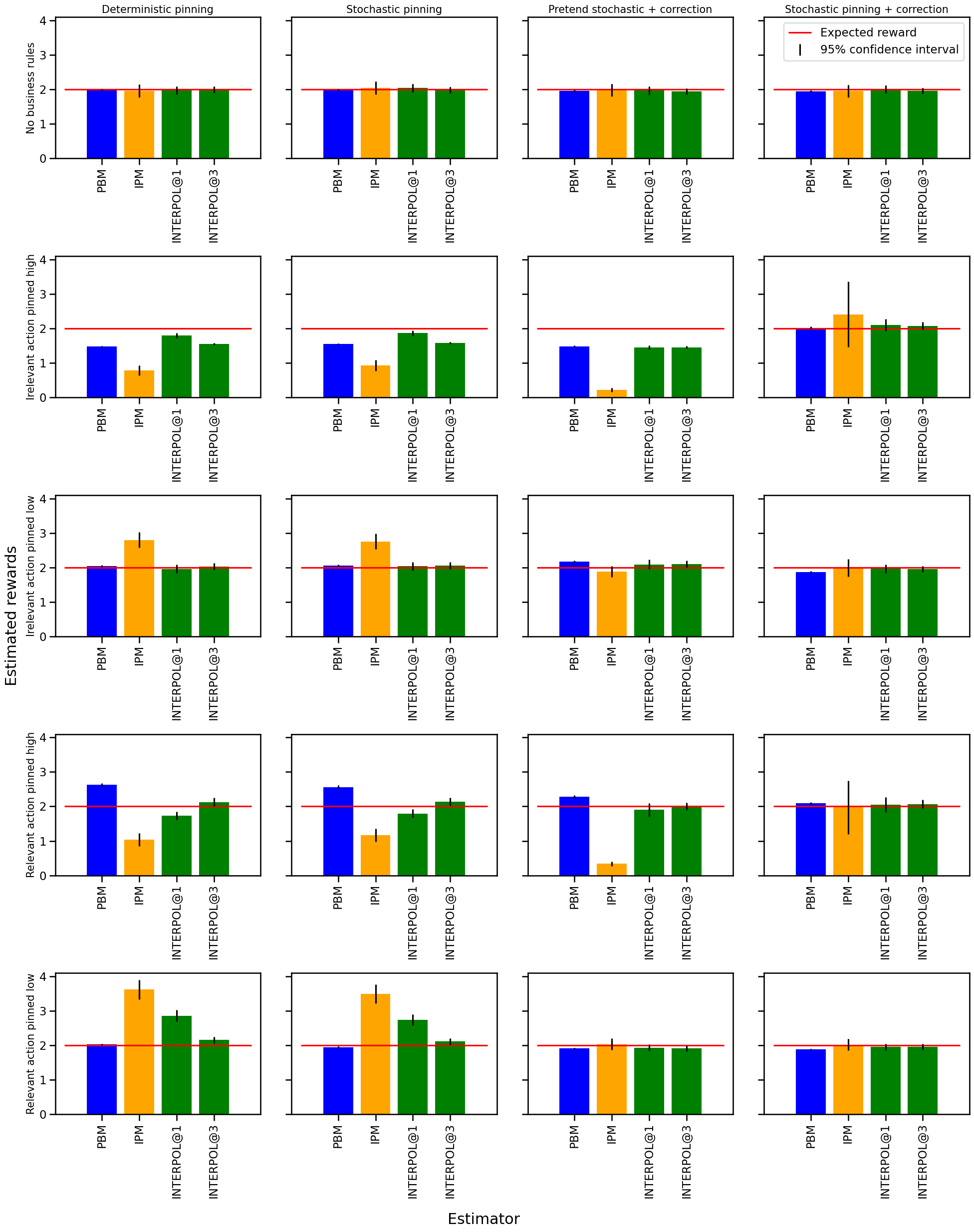}
    \caption{Estimates for additional settings of pinning.}
    \label{fig:additional-results}
\end{figure}

\begin{figure}
    \centering
    \includegraphics[width=\textwidth]{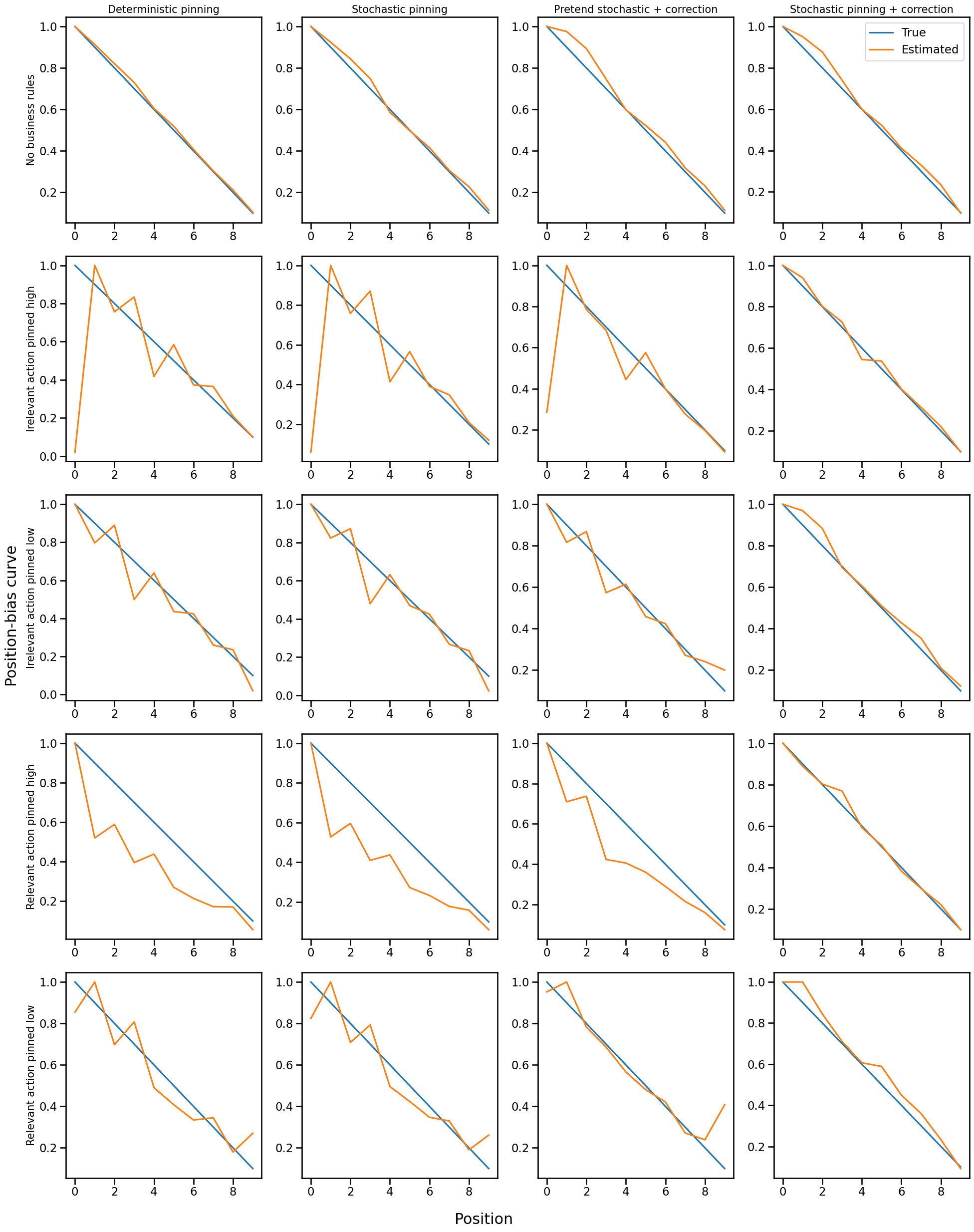}
    \caption{Estimated position-bias curves corresponding to results in figure \ref{fig:additional-results}.}
    \label{fig:additional-curves}
\end{figure}

We analyzed several additional settings of pinning (Figure \ref{fig:additional-results}) and provide corresponding fitted position-bias curves in Figure \ref{fig:additional-curves}:

\begin{itemize}
    \item Item with low/high average relevance pinned to the first/last position (rows in Figures \ref{fig:additional-results} and \ref{fig:additional-curves})
    \item Columns correspond to:
    \begin{itemize}
        \item Pinning applied with 100\% probability with no correction of propensity matrix
        \item Pinning applied with 95\% probability with no correction of propensity matrix
        \item Pinning applied with 100\%, correction was made assuming that pinning was applied with 95\% probability
        \item Pinning applied with 95\%, correction was made assuming that pinning was applied with 95\% probability
    \end{itemize}
\end{itemize}

We can see that assuming stochastic pinning when the actual pinning was applied with 100\% probability often leads to biased estimates, highlighting the importance of having only stochastic business rules. We can also see that INTERPOL estimator is more robust than PBM estimator, as PBM estimator sometimes provides slightly biased results with high confidence. This is caused by the fact that PBM estimator is sensitive to precise estimation of position-bias curve, which is impossible due to inherent variance in the estimation of position-bias curve.

\end{document}